\documentclass[journal]{IEEEtran}
\usepackage{setspace}
\usepackage{cite}
\usepackage{amsmath,amssymb,amsfonts}
\usepackage{graphicx}
\usepackage{algorithm}
\usepackage{caption}
\usepackage{subcaption}
\usepackage{hyperref}
\usepackage{CJK}
\usepackage{algorithmicx}
\usepackage{algpseudocode}
\usepackage{float}
\usepackage{multirow}
\floatname{algorithm}{Algorithm}

\usepackage{geometry}
\hypersetup{
colorlinks=true,
linkcolor=black,
citecolor=black
}
\ifCLASSINFOpdf
\else
\fi
\begin{document}
%
\title{Advanced Codebook Design for SCMA-aided NTNs With Randomly Distributed Users}
%
%
%

\author{Tianyang Hu,
        Qu Luo,~\IEEEmembership{Member,~IEEE,}
        Lixia Xiao, ~\IEEEmembership{Member,~IEEE,}
        Jiaxi Zhou,\\
        Pei Xiao, ~\IEEEmembership{Senior Member,~IEEE,}
        Tao Jiang, ~\IEEEmembership{Fellow,~IEEE.}
\thanks{T. Hu, L. Xiao, J. Zhou, and T. Jiang are with the Research Center of 6G Mobile Communications, the School of Cyber Science and Engineering, Huazhong University of Science and Technology, Wuhan 430074, China, email: \{tianyanghu, lixiaxiao, zhoujiaxi\}@hust.edu.cn; tao.jiang@ieee.org.

Q. Luo and P. Xiao are with 5GIC \& 6GIC, Institute for Communication Systems (ICS), University of Surrey, Guildford GU2 7XH, United Kingdom, e-mail:\{q.u.luo, p.xiao\}@surrey.ac.uk.}}

\maketitle

\begin{abstract}
In this letter, a novel class of sparse codebooks is proposed for sparse code multiple access (SCMA) aided non-terrestrial networks (NTN) with randomly distributed users characterized by Rician fading channels.  
Specifically, we first exploit the upper bound of bit error probability (BEP) of an SCMA-aided NTN with large-scale fading of different users under Rician fading channels. Then, the codebook is designed by employing pulse-amplitude modulation constellation, user-specific rotation and power factors. To further reduce the optimization complexity while maintaining the power diversity of different users, an orthogonal layer-assisted joint layer and power assignment strategy is proposed. Finally, unlike existing SCMA codebook designs that treat all users as one super-user, we propose to minimize the BEP of the worst user to ensure user fairness. The simulation results show that the proposed scheme is capable of providing a substantial performance gain over conventional codebooks.
 
\end{abstract}


%
\IEEEpeerreviewmaketitle

\section{Introduction}
%
%
%
%

\IEEEPARstart {T}{he} next-generation communication systems are expected to support ubiquitous connectivity, accommodate a large number of users,  and facilitate high-spectrum communication. With the widespread proliferation of
Internet-of-Things (IoT) across every corner of this globe, current terrestrial networks are facing challenges in meeting these stringent demands, prompting consideration of non-terrestrial network (NTN)-assisted IoT infrastructure as a complementary solution to extend global coverage \cite{low-projection}. However, it is difficult to achieve massive connectivity for the concurrent orthogonal multiple access-based communications, as the number of accommodated users is limited by the orthogonal resources. Sparse code multiple access (SCMA), as a representative code-domain non-orthogonal multiple access scheme, has emerged as a promising solution to provide high spectrum efficiency communication and massive connectivity in NTNs\cite{NOMA}.

In SCMA, each user's instantaneous input message bits are directly mapped to a multi-dimensional codeword extracted from a meticulously designed sparse codebook. The error rate performance of SCMA relies heavily on the sparse codebooks \cite{11}. By analyzing the pair-wise error probability (PEP) over Gaussian and Rayleigh fading channels,
it is desirable to maximize the minimum Euclidean distance
(MED) and minimum product distance (MPD) of a mother constellation (MC) or a codebook \cite{4}. Following this sprint, the golden angle modulation (GAM) codebook with low peak-to-average power ratio and large MED was proposed in \cite{1}.  Star-quadrature amplitude modulation (star-QAM) was proposed as the MC in \cite{2} for downlink Rayleigh fading channels. In \cite{PIB} a power-imbalanced codebook was proposed by maximizing the MED of the superimposed constellation while keeping the MPD larger than a threshold. In addition, a class of low-projection codebooks have been proposed  to achieve low-complexity detection \cite{low-projection}.


It should be noted that the existing SCMA codebooks for downlink channels were mostly optimized by treating all SCMA users as one super-user.
 Generally speaking, these works implicitly assumed that all SCMA users have similar locations and channel conditions \cite{3,7,8,9,SSDSCMA},    with very few considerations on the codebook design for randomly distributed users within a cell.
 Due to path loss discrepancies, users at different locations may experience different error rate performances, resulting in user unfairness.
  Moreover, most SCMA codebook designs focus on Gaussian and Rayleigh fading channels,   while the channels of NTNs are typically modeled as Rician fading. Against this background, our primary objective is to design advanced codebooks for SCMA-aided NTNs with randomly distributed users. 
  
The main contributions are summarized as follows:
{
\begin{itemize}
\item We use order statistics to analyze the system performance of an SCMA-aided NTN with user location variability and large-scale fading. With the derived pair-wise error probability of different users, novel codebook design metrics have been proposed to enhance the system performance and user fairness.
\item We propose an orthogonal layer-assisted joint layer and power assignment strategy aimed at reducing the complexity of optimization through group-based parameter sharing, while preserving the power diversity among different groups of users. 
 \end{itemize}}

\section{SCMA system model} \label{System model}
\begin{figure}
\hspace{100em}
    \centering  
    \includegraphics[width=0.7\linewidth]{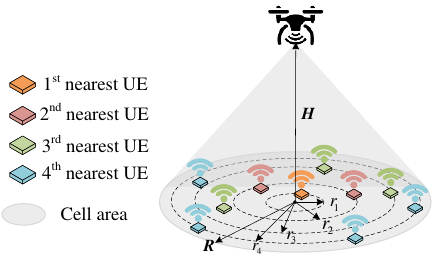}
    \caption{An example of  NTN network with randomly distributed users.}
    \label{fig:system model}
    \vspace{-1em}
\end{figure}

\begin{figure}
    \centering  
\includegraphics[width=0.7\linewidth]{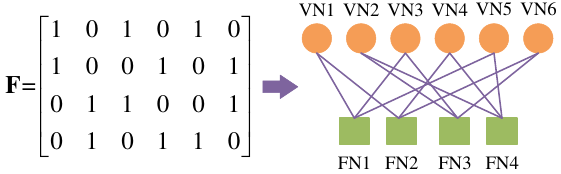}
    \caption{An illustration of factor graph in an SCMA system.}
    \label{fig:factor graph}
    \vspace{-1.5em}
\end{figure}

We consider an downlink SCMA-aided NTN system with an air base station (ABS) located at the center, serving $J$ users,  as shown in Fig.\ref{fig:system model}. Assume that the cell is a circular area of radius $R$ and that users are randomly distributed in the cell.  In a downlink SCMA-aided NTN system, the ABS transmits the messages to $J$ users through orthogonal resource nodes (RNs) of $K$, where the overloading factor is defined as $\lambda=J/K>1$. Each user is assigned a unique codebook, denoted by $ \boldsymbol{\mathcal { X }}_{j} \in \mathbb C^{K \times M}$, consisting of $M$ codewords with a dimension of $K$.  Each user's incoming message bits are assigned to a specific transmitted codeword, with the assignment procedure for the $j$th user described as 
\begin{equation}
\small
    f_j:\mathbb{B}^{\log_2M}  \rightarrow \boldsymbol{\mathcal {X}}_{j},  \text{i.e., } \mathbf {x}_{j} =     f_j(\mathbf b_{j}),
\end{equation}
where $\mathbf b_{j} \in \mathbb B^{\log_2M \times 1}$ is the input bit message of user $j$ and $\mathbf {x}_{j} \in \boldsymbol{\mathcal {X}}_{j} $ is the transmitted codeword.  
Each codeword only contains $N $ nonzero entities with $N<K$, and the nonzero positions remain the same within each codebook. The arrangement of the $J$ SCMA codebooks is depicted by the indicator matrix $\mathbf {F}_{K \times J} = [ \mathbf {f}_1, \ldots, \mathbf {f}_J ] \in \mathbb {B}^{K\times J}$. The component in $\mathbf{F}_{K \times J}$ is denoted as $f_{k,j}$, and a connection between the variable node $j$ and resource node $k$ exists exclusively when $f_{k,j}=1$.   Fig.\ref{fig:factor graph} illustrates an example of an indicator matrix and the associated factor graph with $K=4,J=6,N=2$.


Denoted $\mathbf {h}_{j}=[ {h}_{j,1}, {h}_{j,2}\ldots , {h}_{j,K}]^{{ \mathcal T}}$ by the channel between the ABS and the user $j$ with each component $ {h}_{j,k}= {g}_{j,k} d_{j}^{-\frac {\alpha }{2}}$.    ${g}_{j,k}$ is modeled as Rician fading channel, i.e., ${g}_{j,k}\sim\mathcal{C}\mathcal{N}\left( \sqrt{\frac{\kappa }{1+\kappa }},\sqrt{\frac{1}{1+\kappa }} \right)$, where $\kappa$ denotes the Rician factor,    $d_j$ is the distance between the ABS and the $j$th user  and $\alpha \geq 1$ is the path loss coefficient.   
{Define the ratio of the ABS altitude to the cell radius as $c_1=H/R$, and the ratio of the distance between the ABS nadir point and user $j$ to the cell radius as $c_{2,j}=r_j/R$. Then the distance between the ABS and the user $j$ can be expressed as $d_{j}^{-\alpha/2 }=R^{-\alpha/2}{(c_1^2+c_{2,j}^2)}^{-\alpha /4}$. If the path loss experienced at a signal transmission distance $R$ is the fundamental path loss, after pre-compensating all users' fundamental path losses, the final path loss of user $j$ is given by $\hat{d}_j^{-\alpha/2}=(c_1^2+c_{2,j}^2)^{-\alpha/4}$.}
In downlink SCMA systems, the ABS transmits the superimposed codewords to $J$ users and the user $j$ receives the signal as
\begin{equation} 
\small
    {{\mathbf{y}}_{j}}=(c_1^2+c_{2,j}^2)^{-\alpha/4}\text{diag}(\mathbf{g}_j)\sum\limits_{l=1}^{J}{\sqrt{{{p}_{l}}}{{\mathbf{x}}_{l}}}+\mathbf{n}_j,
\end{equation}
where   ${{p}_{l}}$ denotes the allocated power to user $l$, and $\mathbf{n}_j$ is the Gaussian white noise with $ \mathcal{C}\mathcal{N}(0,{{N}_{0}})$ entries. 

\textit{Remark 1:
{Considering the movement of the ABS, the optimal codebook depends on the instantaneous distances between the ABS and users. In this case, the codebook design becomes an online process, which is ineffective due to the significant computational complexity of codebook design. In this letter, we propose an offline codebook design scheme.  Specifically, we assume that $J$ users are randomly distributed within the cell, and let the ordered distances given by   $d_1  \leq  d_2\leq,  \ldots, \leq  d_J $. Then, the codebook is designed by averaging over the statistic distribution of $d_j$, allowing the designed codebook to be directly applied in online deployment. In this scenario, the ABS does not require precise knowledge of absolute distances to individual users; instead, only the relative proximity ordering among users needs to be determined.} }

\vspace{-0.2em}
\section{Performance Analysis} \label{CB design}
\label{Perfor}
{Define the transmitted  codewords of $J$ users as $\mathbf{{X}}=[\mathbf{{x}}_{1},\mathbf{{x}}_{2},...,\mathbf{{x}}_{J}]$.  Meanwhile, the erroneously decoded codewords of user $j$ is denoted by
$\mathbf{\hat{X}}_{j}=[\mathbf{\hat{x}}_{1},\mathbf{\hat{x}}_{2},...,\mathbf{\hat{x}}_{J}]$, where $\mathbf{\hat{x}}_{j} \in \boldsymbol{\mathcal { X }}_{j}, \mathbf{\hat{X}}_{j} \neq \mathbf{X}$.}
To proceed, let $\text{Pr}\{{{\mathbf{X}} }\to {{\mathbf{\hat{X}}}_{j}}\}$ be the PEP of ${{\mathbf{X}} } $ and $ {{\mathbf{\hat{X}}}_{j}}$, which is conditioned on $\mathbf h_j $. Then, according to \cite{4}, the PEP of the $j$th user is given by
\begin{equation} \label{PEP origin}
\setlength\belowdisplayskip{0pt}
\small
    \text{Pr}\{{{\mathbf{X}} }\to {{\mathbf{\hat{X}}}_{j}}\} \approx {{\mathbb{E}}_{{{\mathbf{h}}_{1}},\cdots ,{{\mathbf{h}}_{K}}}}\left[ Q\left( \sqrt{\frac{d({{\mathbf{X}} },{{{\mathbf{\hat{X}}}}_{j}})}{2{{N}_{0}}}} \right) \right],
\end{equation}
where 
\begin{equation} \label{codeword distance} 
\small
    d({{\mathbf{X}}},{{\mathbf{\hat{X}}}_{j}})={{\left\|(c_1^2+c_{2,j}^2)^{-\frac{\alpha}{2}}\text{diag}(\mathbf{g}_j)\sum\limits_{l=1}^{J}{\sqrt{{{p}_{l}}}({{\mathbf{x}}_{l}}-{{\widehat{\mathbf{x}}}_{l}})} \right\|}^{2}}.
 \end{equation}

{Let ${{\beta }_{k}}={(c_1^2+c_{2,j}^2)}^{-\frac{\alpha}{2}}{\left\| \sum\limits_{l\in {{\Phi}_{k}}}{\sqrt{{{p}_{l}}}({{x}_{l}}[k]-{{\widehat{x}}_{l}}[k])} \right\|}$,  where ${{\Phi}_{k}}$ denotes the set of users colliding over the $k$th RN. Since the $j$th user's position relative to the ABS nadir point is considered unknown, the PEP of the $j$th user conditioned on distance ratio $c_{2,j}$  is given as}
\begin{equation}
\small
\text{Pr}\{{{\mathbf{X}}}\to {{\mathbf{\hat{X}}}_{j}}|c_{2,j}\}={{\mathbb{E}}_{{{g }_{j,1}},\cdots ,{{g }_{j,K}}}}\left[ Q\left( \sqrt{\frac{1}{2{{N}_{0}}}\sum\limits_{k=1}^{K}g_{j,k}^2{ {{\left| {{\beta }_{k}} \right|}^{2}}}} \right) \right].
\label{UE_PEP}
\end{equation}
{A tight approximation of $Q$-function is given by $Q(x)\approx \frac{1}{12}{{e}^{-{{x}^{2}}/2}}+\frac{1}{4}{{e}^{-2{{x}^{2}}/3}}$ \cite{low-projection}.  In addition, the   distribution of $g_{j,k}^2$ belongs to the noncentral chi-square distribution whose moment-generating function is $M_{|g_k|^2} (s) = \frac{1+\kappa}{1 +\kappa+ s}\exp \left(- \frac{\kappa s}{1 +\kappa + s} \right)$. By substituting \eqref{UE_PEP} into the above approximation and  based on the  $M_{|h_k |^2} (s) $, one has}
\begin{equation} 
\setlength\belowdisplayskip{0pt}
\small
\begin{split}
    \text{Pr}\{{{\mathbf{X}}}\to {{\mathbf{\hat{X}}}_{j}}|c_{2,j}\} 
    &\approx \frac{1}{12} \prod\limits_{k=1}^{K}{\frac{1+{{\kappa }}}{1+{{\kappa }}+{s_k/4}}\exp \left( -\frac{{{\kappa }}{s_k/4}}{1+{{\kappa }}+{s_k/4}} \right)}+ \\
    &\frac{1}{4} \prod\limits_{k=1}^{K}{\frac{1+{{\kappa }}}{1+{{\kappa }}+{s_k/3}}\exp \left( -\frac{{{\kappa }}{s_k/3}}{1+{{\kappa }}+{s_k/3}} \right)},
   \label{fin PEP}
\end{split}
\end{equation}
where
\begin{equation}
\small 
\label{sk}
\begin{split}
    {s_k}&=\frac{{\big| \sum\nolimits_{l\in {{\Phi}_{k}}}{\sqrt{{{p}_{l}}}({{x}_{l}}[k]-{{\widehat{x}}_{l}}[k])} \big|}^{2}}{{N}_{0}{(c_1^2+c_{2,j}^2)}^{\alpha /2} R^{\alpha}}.
\end{split}
\end{equation}



For a polar coordinate system constructed with the ABS nadir point as the origin, the position of the user $j$ can be characterized by the tensor angle $\theta_j\sim  \mathcal{U}\left( 0, 2\pi  \right)$ and the distance $r_j$ from the user to the center. The probability density function (PDF) and the cumulative distribution function (CDF) of $r$ can be respectively expressed as
\begin{equation}
\small
    f(r_j)=\frac{2}{{{R}^{2}}}r_j,F(r_j)=\frac{1}{{{R}^{2}}}{{r_j}^{2}}.  \label{pdf+cdf}
\end{equation}
Upon sorting the users from nearest to farthest, the distances from the center to the $J$ users form an independent and identically distributed random sequence $d_1  \leq  d_2\leq,  \ldots, \leq  d_J $. According to the order  statistic theory \cite{order-sta}, the PDF $f_{(j)}$ of the ordered $d_j$ is given by
\begin{equation}
\small
{{f}_{(j)}}(x)=Jf(x)\left( \begin{matrix}
   J-1  \\
   j-1  \\
\end{matrix} \right)F{{(x)}^{j-1}}{{(1-F(x))}^{J-j}}.  \label{ODST pdf}
\end{equation}
Substituting \eqref{pdf+cdf} into \eqref{ODST pdf} and letting $x={{r_j}^{2}}\text{/}{{R}^{2}}$, the average values of the distance ratio of the $j$th user is given by
\begin{equation}
\label{expectation}
\small
\begin{split}
\mathbb E(c_{2,j})&=\frac{1}{R}\int_{\text{0}}^{\text{1}}{xJ\left( \begin{matrix}
   J-1  \\
   j-1  \\
\end{matrix} \right){{x}^{j-1}}{{(1-x)}^{J-j}}\frac{R}{\sqrt{x}}dx} \\ 
 &=\frac{1}{B(a,b)}\int_{0}^{1}{{{x}^{a-1/2}}{{(1-x)}^{b-1}}dx} \\ 
 &=\frac{\Gamma (a+1/2)\Gamma (a+b)}{\Gamma (a+b+1/2)\Gamma (a)},  
\end{split}
\end{equation}
where $B(a,b)$ represents the Beta distribution with $a=j,b=J-j+1$. By substituting the average of $c_{2,j}$ into \eqref{sk}, the unconditional $\text{Pr}\{\mathbf{X}\to \mathbf{\hat{X}}_j\}$ is obtained. It is important to emphasize that, for the  BER of each user, only the cases where the information intended for that particular user is decoded incorrectly should be taken into account. Namely, the upper bound of BEP of user $j$ is expressed as
\begin{equation} \label{BEP}
\small
\setlength\belowdisplayskip{0pt}
\begin{split}
    P_{\text{e},j} &\approx  \frac{1}{{{M}^{J}}{{\log }_{2}}(M)}\underset{\mathbf{X}}{\mathop{\sum }}\,\underset{\mathbf{\hat{X}},{{\mathbf x}_{j}}\ne {{{\hat{\mathbf x}}}_{j}}}{\mathop{\sum }}\,n\left( {{\mathbf x}_{j}},{{{\hat{ \mathbf x}}}_{j}} \right)\text{Pr}\{\mathbf{X}\to \mathbf{\hat{X}}_j\},
\end{split}
\end{equation}
where $n({{\mathbf x}_{j}},{{\hat{\mathbf x}}_{j}})$ stands for the number of error bits of user $j$. The average and worst BEPs of $J$ users are given by  
\begin{equation} 
\setlength\abovedisplayskip{0pt}
\setlength\belowdisplayskip{0pt}
\small
\begin{split}
    P_{\text{e}, \text{ave}}\approx  \frac{1}{J} \sum_{j=1}^{J}  P_{\text{e},j},   \quad   P_{\text{e}, \text{wor}}\approx   \max_{1\leq j \leq J} P_{\text{e},j}. 
\end{split}
\end{equation}

\section{Codebook construction}
The proposed approach mainly involves three steps: 1) MC design;   2) multiuser codebook construction and 3) parameter optimization.

\textit{Step 1: Generate the MC.}  The  pulse-amplitude modulation is employed to construct the mother constellation  owing to its characteristic of  symbolic energy multiplicity \cite{2}. With interleaving and permutation, the elements in the  $n$th-dimension of the MC $\boldsymbol {\mathcal A}$ are generated by
\begin{equation} \label{MC}
\small
    {\mathbf{A}^{M}_{n}}=\left\{ \begin{matrix}
   \left[ -{a}_\frac{M}{2},-a_{\frac{M}{2}-1}\ldots ,-a_{1},a_{1},\ldots ,{a}_\frac{M}{2} \right],\text{ if n is odd}  \\
   \left[ -a_1,a_\frac{M}{2},-a_2,a_{\frac{M}{2}-1},\ldots,-a_\frac{M}{2},a_1 \right],\text{ if n is even}  \\
\end{matrix} \right.
\end{equation}
where ${{a}_{m}}=(m(\delta -1)+(2-\delta)),m=1,2,\ldots ,M/2$ and $\delta \in \left( \text{1,}\text{+}\infty  \right)$  denotes the amplitude parameter to be optimized  later. The energy of each dimension of the codebook is
\begin{equation} \label{energy}
\small
    E\text{=}M(2-\delta )(1+\frac{M}{2}\delta -\frac{M}{2})+\frac{M(M+2)(M+1){{(\delta -1)}^{2}}}{12}.
\end{equation}

\textit{Step 2: Generate  sparse codebooks for different users.}  Upon obtaining the MC, the codebook for the $j$th user can be constructed by
\begin{equation}
\small
    {\boldsymbol{\mathcal X}_j}^{K\times M}= \sqrt{M/(NE)}\mathbf{V}_{j}^{K\times N}\mathbf{\Delta }_{j}^{N\times N}{{\boldsymbol {\mathcal A}}^{N\times M}}, 
\end{equation}
where $\mathbf{V}_j \in \mathbb B^{K \times N}$ denotes the mapping matrix that converts the $N$-dimensional dense constellation into a $K$-dimensional  sparse codebook and $\sqrt{{M} / (NE)}$  serves as the energy normalization factor to guarantee $\text{Tr}(  {\boldsymbol{\mathcal X}_j}^{H}   {\boldsymbol{\mathcal X}_j} )=M$.      $\mathbf {V}_{j}$  can be constructed based on the position of the zero elements of ${{\mathbf{f}}_{j}}$ by inserting all-zero row vectors into the identity matrix ${{\mathbf{I}}_{N}}$. For instance,  in the case of $\mathbf F$ depicted in Fig.  \ref{fig:factor graph}, we have ${{\mathbf{V}}_{1}}=\left[ \begin{matrix}
  0 & 1  & 0 & 0  \\
  0 & 0 &  0 & 1  \\
\end{matrix} \right]^{ T}$ and   ${{\mathbf{V}}_{2}}=\left[ \begin{matrix}
  1 & 0  & 0 & 0  \\
  0 & 0 &  1 & 0  \\
\end{matrix} \right]^{ T}$,   and $\mathbf {V}_{j}, j= 3,\ldots,6$ can be generated in the same way.       $\mathbf{\Delta}_j= \text{diag} (e^{i{{\theta }_{j,1}}},...,e^{i{{\theta }_{j,N}}} )$,  where $0<{\theta }_{j,n}<\pi,\forall n=1,\ldots,N$, is the rotation matrix identifying the $j$th user amongst those sharing the same RE.

{\textit{Step 3: Codebook optimization}.  Considering user fairness, we propose to minimize the worst BER of the $J$ users. Namely, the codebook design can be formulated as}
\begin{subequations} 
\label{Ometric} 
\small
\begin{align}
      [p_{1,\cdots,J},\mathcal X_{1,\cdots,J}] = &\arg \underset{\delta,p_j,\mathbf{V}_j,\mathbf{\Delta}_j}\min \underset{1\le j\le J}\max {{P}_{e,j}}(p_j{\boldsymbol{\mathcal X}_j}) \tag{\ref{Ometric}}\\ 
     \text{ s.t. } \text{ } &{\boldsymbol{\mathcal X}_j} = \sqrt{M/(NE)}\mathbf{V}_{j}\mathbf{\Delta }_{j}\boldsymbol{\mathcal A}, \delta>1, \label{Ometric a}\\    
      &\sum\nolimits_{j=1}^{J}{{p}_{j}} = J,  p_j>0, j=1,\ldots, J,  \label{Ometric b}\\
      &0<{\theta }_{j,n}<\pi,\forall n=1,\ldots,N. \label{Ometric c} 
 \end{align}
\end{subequations}
\vspace{-1.5em}

{Given that each RE is shared among $d_f$ users, a minimum of $d_f$ distinct phase configurations are required to ensure user separability. Consequently, the system involves $J+d_f +1$ design parameters: $d_f$ phase rotation angles, $J$  power parameters, and the amplitude parameter $\delta$ in $\boldsymbol{\mathcal A}$.  However, the parameter count scales proportionally with the size of the factor graph, leading to higher computational complexity for codebook design. Moreover, when the widely employed heuristic algorithms are used for optimization, the performance of the codebook may deteriorate.  
These challenges highlight the need to reduce the number of parameters, especially in larger factor graphs, to alleviate computational burdens and improve codebook optimization efficiency. } 

{
 Note that the power allocation can be merged with the process of codebook design, which means the power assigned to each user is combined with the constellation points in their corresponding codebook. Additionally, it is critical to balance power across different RNs as if an RN operates with insufficient power, the distance between distinct codewords at that RN will be compromised. At this point, we adopt $d_f$ power factors and assume uniform power allocation across all RNs. Denoted ${{q}_{i}}={\rho}_{i} e^{j{{\theta }_{i}}}$,  $1\leq i \leq d_f$, by the constellation operator combining the power scaling  $0<{\rho}_{i}<1$ and phase rotation angle $0\leq{\theta }_{i}<\pi$.   The problem can be formulated as the allocation of  $q_i$ into the non-zero entry of the factor graph. Assume the $j$th SCMA layer, i.e.  $\mathbf f_j$ is assigned to user $j$. Subsequently, the allocation of $q_i$ adheres to these guidelines:
\begin{itemize}
\item It is desirable to assign a large power to the user that has higher propagation losses  for user fairness. 
\item  Each RN has the same power factors to achieve the power balance among different RNs.
\end{itemize}
}
{To address the above allocation, a joint layer and power allocation method assisted by orthogonal layers is proposed based on greedy algorithm. The key idea involves organizing SCMA layers into $T$ orthogonal groups, labeled as $\mathcal{G}_t$ for the $t$th group, where the SCMA layers are orthogonal to each other, i.e., $\mathbf f_{j}^{T}{\mathbf{f}_{j'}} = \mathbf{0}$ for all $j' \neq j$ in $\mathcal{G}_t$. Then, each group is assigned with the same $q_i$ to satisfy the power balance among different RNs. And the final obtained signature mathix is defined as $\boldsymbol{\mathcal S}$.}

The detailed algorithm is summarized in Algorithm \ref{ORA}.
For ease of allocation, both users and $q_i$ are arranged in ascending order according to $d_i$ and $\rho_i$ respectively.  The function “$\text{find}(\cdot)$”    returns a vector containing the index of each non-zero element of the input. {The results corresponding to the systems of $\lambda=150\%$ and $\lambda=200\%$ are presented as follows.}
\begin{equation}
\label{f46}
\small
   {\mathbf S_{4\times6}} \text{=}\left[ \begin{matrix}
   q_1 & 0 & q_2 & 0 & 0 & q_3  \\
   0 & q_1 & q_2 & 0 & q_3 & 0  \\
   q_1 & 0 & 0 & q_2 & q_3 & 0  \\
   0 & q_1 & 0 & q_2 & 0 & q_3  \\
\end{matrix} \right],
\end{equation}
\begin{equation}
\label{f510}
\small
   {\mathbf S_{5\times10}} \text{=}\left[ \begin{matrix}
   q_1 & 0 & 0 & q_2 & 0 & 0 & q_3 & 0 & 0 & q_4\\
   q_1 & 0 & q_2 & 0 & 0 & q_3 & 0 & q_4 & 0 & 0\\
   0 & q_1 & 0 & q_2 & 0 & q_3 & 0 & 0 & q_4 & 0\\
   0 & q_1 & 0 & 0 & q_2 & 0 & 0 & q_3 & 0 & q_4\\
   0 & 0 & q_1 & 0 & q_2 & 0 & q_3 & 0 & q_4 & 0\\
\end{matrix} \right].
\end{equation}

In the ${\mathbf S_{4\times6}}$, SCMA layers are divided into $d_f=3$ groups, each containing two layers. Each group employs identical phase and power. For the ${\mathbf S_{5\times10}}$, we have $\mathcal{G}_1 = \{1, 2\}$, $\mathcal{G}_2 = \{4, 5\}$, $\mathcal{G}_3 = \{6, 7\}$, and $\mathcal{G}_4 = \{9, 10\}$. Layer $3$ and layer $8$ are assigned $q_2, q_1$ and $q_4, q_3$ to satisfy the above rules.

Finally,  the codebooks can be obtained by solving  (\ref{Ometric}) with the the power and phase operators   given in (\ref{f46}) and (\ref{f510}).   Similar to existing SCMA codebook design  works, (\ref{Ometric}) is tackled using a numerical global search approach via the MATLAB Global Optimization Toolbox. Specifically, the genetic algorithm (GA) is applied to solve the optimization problem, which starts with an initial population and iteratively selects individuals based on their fitness.    In the conducted simulations, the population size is configured at $50$ and the maximum number of iterations is limited  to $20$. {The SNR used in calculating the design metric is set to $12$ dB for the codebook of $\lambda=150\%$ and $15$ dB for the codebook of $\lambda=200\%$.}
Despite the varied resultant codebooks, they share similar theoretical PEP values and exhibit superior BER performances in simulations on both system performance and user fairness. This improvement comes with a slight increase in complexity, but since the optimization is done offline, the additional complexity has minimal impact and allows for direct deployment on onboard equipment.

\begin{algorithm} [t]
\small
\caption{Joint layer and power assignment}
\label{ORA}
\begin{algorithmic}[1]
\Require $[q_1,\ldots,q_{d_f}],K,J,N,d_f$.
\Ensure $\boldsymbol{\mathcal S}$.

\State Set $L=\lfloor K/N \rfloor$ and $R=J-Ld_f$.
\State Initialize $\boldsymbol{\mathcal Q}=[0]_{K\times L\times d_f}$, $\boldsymbol{\mathcal F}=[0]_{K\times d_f}$, $\boldsymbol{\mathcal R}=[\text{ }]$.
\State Generate all the possible combinations of $N$ ones and $K-N$ zeros, named as $\mathbf{P}$. Set $\boldsymbol{\mathcal I}=[d_f,\ldots,d_f]_{1\times K}$.

\For{$u=1:d_f$}
\For{$l=1:L$}
\State Find one column $\mathbf{p}\in \mathbf{P}$ that ${\mathbf{p}}^T\cdot \boldsymbol{\mathcal Q}(:,:,u)=\mathbf{0}$. 
\If{$\mathbf{p}$ does not exist}
\State $R=R-1$, $\boldsymbol{\mathcal R}=\boldsymbol{\mathcal R} \cup \boldsymbol{\mathcal Q}(:,l-1,u)$.
\State $\boldsymbol{\mathcal Q}(:,l-1,u)=\mathbf{0}$, $l=l-2$. \textbf{Continue}.
\EndIf
\State $\boldsymbol{\mathcal Q}(:,l,u)=\mathbf{p}$, $\boldsymbol{\mathcal F}(:,u)=\boldsymbol{\mathcal F}(:,u)+\mathbf{p}$, $\mathbf{P}=\mathbf{P}\backslash \mathbf{p}$.
\EndFor
\EndFor

\If{$R\sim=0$}
\State Fetch $R$ columns $\{\mathbf{p}\}_{K\times R} \in \mathbf{P}$, $\boldsymbol{\mathcal R}=\boldsymbol{\mathcal R} \cup \{\mathbf{p}\}_{K\times R}$.
\EndIf

\For{$v=1:(J-Ld_f)$}
\State Set $\mathbf{n}'=\text{find}(\boldsymbol{\mathcal R}(:,v))$.
\For{$j=1:N$}
\State Find the $g$th layer of $\boldsymbol{\mathcal{Q}}$ that $\boldsymbol{\mathcal Q}(\mathbf{n}'(j),:,g)=0$.
\State Set $\mathbf{I}=\boldsymbol{\mathcal I}(\text{find}(\boldsymbol{\mathcal F}(:,g)))$.
\State Replace the ones in $\boldsymbol{\mathcal Q}(:,:,g)$ with $q_x$, where $x$ stands for every element in $\mathbf{I}$. Set $\boldsymbol{\mathcal I}(\text{find}(\boldsymbol{\mathcal F}(:,g)))=\mathbf{I}-1$.
\State $\boldsymbol{\mathcal R}(\mathbf{n}'(j),v)=q_{\boldsymbol{\mathcal I}(\mathbf{n}'(j))}$, $\boldsymbol{\mathcal I}(\mathbf{n}'(j))=\boldsymbol{\mathcal I}(\mathbf{n}'(j))-1$.
\EndFor
\EndFor
\State Update $\boldsymbol{\mathcal S}=\boldsymbol{\mathcal Q}\cup\boldsymbol{\mathcal R}$.
\State Sort each column of $\boldsymbol{\mathcal S}$ in ascending order of total power. 
\end{algorithmic}
\end{algorithm}

\vspace{-0.5em}
\section{Simulation results}  \label{simulation}
In this section, numerical results are provided to evaluate the performance of the proposed codebooks. The proposed codebooks are available at our GitHub project \footnote{https://github.com/ethanlq/SCMA-codebook}. In all simulations, the downlink Rician channel with perfect channel state information is assumed. For simplicity, we set $c_1=1$ according to \cite{NTN} and employ 4-ary codebooks. Meanwhile, it is assumed that the users are evenly distributed in the cell and that the fundamental path losses for all users in the cell have been previously equalized.  Additionally, the path loss factor is set to $\alpha=3$. In each Monte Carlo simulation, the distances of $J$ users  within the cell range are generated uniformly  and sorted in descending order, which can be regarded as the codebook allocation process. The final BER result at each SNR value is obtained by averaging over a large number of Monte Carlo simulations.

\begin{figure}
    \centering
    \includegraphics[width=0.62\linewidth]{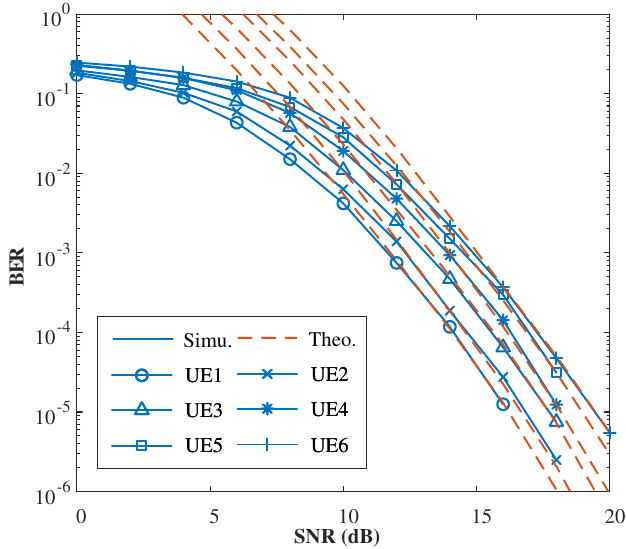}
    \caption{Analytical  and simulated BERs of different users with the GAM codebook at $\kappa=10$.}
    \label{fig:PEP}
\vspace{-1.5em}
\end{figure}

Fig.\ref{fig:PEP} presents the simulated and analytical BERs with the GAM codebook in \cite{1} on $\kappa=10$. As shown in Fig.\ref{fig:PEP}, the analytical curves are well-matched to the simulated curves for all six users in the large SNR regions while in the low SNR regions, significant gaps are observed. This is reasonable as the approximation of PEP is employed in calculating \eqref{BEP} and the approximation only holds tight for a small value of $N_0$. Moreover, the differences in BER among users suggest that the current codebooks cannot guarantee user fairness due to varying propagation losses among the $J$ users.

\begin{figure}[htbp]
     \centering
     \begin{subfigure}[b]{0.22\textwidth}
         \includegraphics[width=\textwidth]{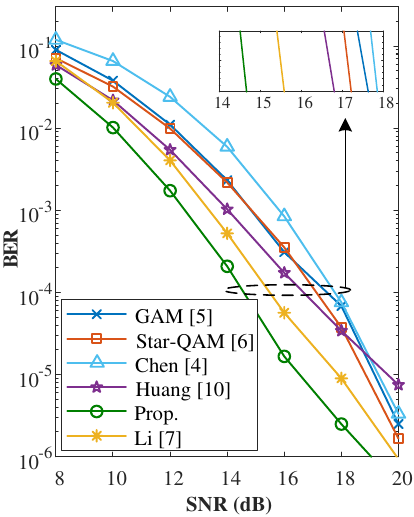}
         \caption{$\kappa=10$}
         \label{fig:CB k=10}
     \end{subfigure}
     \begin{subfigure}[b]{0.22\textwidth}
         \includegraphics[width=\textwidth]{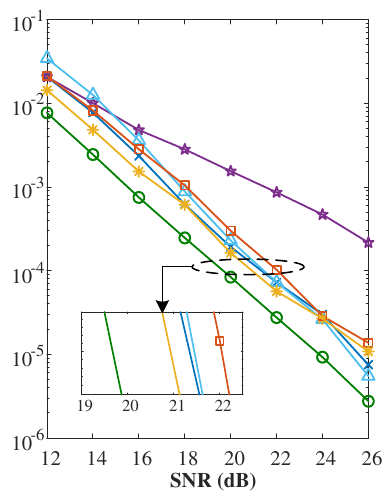}
         \caption{$\kappa=2$}
         \label{fig:CB k=2}
     \end{subfigure}
    \caption{Performance comparison of the proposed and conventional codebooks with $\lambda=150\%$.}
    \label{fig:CB performance}
\end{figure}
\begin{figure}[htbp]
    \centering    \includegraphics[width=0.62\linewidth]{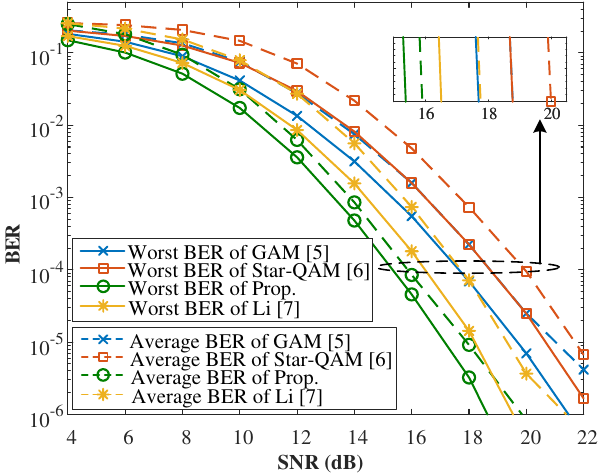}
    \caption{{The average and worst BER performance of the proposed and conventional codebooks with $\lambda=200\%$ and $\kappa=10$.}}
    \label{fig:200CB}
    \vspace{-1.5em}
\end{figure}

{
Fig.\ref{fig:CB performance} compares the worst user BER performance of the proposed codebook with the GAM, Huang\cite{7}, Chen \cite{4}, Li \cite{PIB} and star-QAM \cite{2} codebooks with $\lambda=150\%$ under Rician factors of $10$ and $2$. It should be noted that the Huang and star-QAM codebooks are designed for Gaussian channel while the GAM and Chen codebooks are designed for Rayleigh channel. Meanwhile, the Li codebook is also optimized for user power, providing good performance in both Gaussian and Rayleigh channels.} As shown in Fig.\ref{fig:CB k=10}, the proposed codebook achieves the best BER performance. Notably, the proposed codebook provides approximately $1$ dB gain over the Li codebook and $1.9$ dB over the Huang codebook at BER$=10^{-4}$ for $\kappa=10$. Further observations from Fig.\ref{fig:CB k=2} show that the proposed codebook provides about $1.9$ dB gain over the GAM and Chen codebooks, and approximately $2.2$ dB gain over the star-QAM codebook at BER$=10^{-4}$ for $\kappa=2$.

{
Fig.\ref{fig:200CB} shows the average and worst BER performances of the proposed, GAM, star-QAM and Li codebooks for $\lambda=200\%$ and $\kappa=10$. The proposed codebook outperforms all benchmark designs in both average BER and worst-user BER performance. At BER$=10^{-4}$, the proposed codebook achieves about $1.9$ dB, $2.9$ dB and $4.1$ dB gains over the Li, GAM and star-QAM codebook, respectively, in worst-user BER. Furthermore, the proposed codebook exhibits a smaller performance gap between worst-user BER and average BER. At BER$=10^{-4}$, the gap of the proposed codebook is $0.5$ dB, compared to $1.1$ dB observed in other codebooks. This narrower differential reflects improved fairness while maintaining competitive BER performance.
}

Table \ref{tab:gain} shows the worst BER gain of the proposed codebook over the conventional codebooks at BER$=10^{-4}$ with $\lambda=150\%$ and $\lambda=200\%$ on $\kappa=0,2,5,15,30,100$. The Rayleigh fading and AWGN channels are the special cases of Rician fading channels, corresponding to $\kappa =0$ and $\kappa \rightarrow \infty$, respectively. It can be seen from the table that our proposed codebook achieves the best performance on all the values of $\kappa$ and the gains become larger when the overloading factor increases.

\begin{table}[htbp]
\small
\begin{center}
\caption{The provided gains of the worst BER over other codebooks at BER$=10^{-4}$ with $\lambda=150\%,\text{ } 200\%$.}
\label{tab:gain}
\begin{tabular}{| c | c  c | c  c | c  c |}
\hline
\multirow{2}*{$\kappa$} & \multicolumn{2}{ c }{Li [7]} & \multicolumn{2}{| c }{GAM [5]}& \multicolumn{2}{| c |}{Star-QAM [6]}\\
\multicolumn{1}{|c|}{~} & $150\%$ & $200\%$ & $150\%$ & $200\%$ & $150\%$ & $200\%$\\
\hline
0&0.8dB&2dB&0.6dB&1.5dB&2dB&2.7dB\\
\hline
2&0.8dB&2.7dB&1.9dB&2dB&2.2dB&2.7dB\\
\hline
5&0.5dB&1.8dB&1.5dB&2.3dB&1.7dB&3.1dB\\
\hline
15&0.9dB&1.9dB&3.2dB&3.7dB&2.7dB&4.5dB\\
\hline
30&0.9dB&1.9dB&2.9dB&3.8dB&2.5dB&4.9dB\\
\hline
100&1.1dB&1.9dB&3.2dB&3.6dB&2.1dB&4.5dB\\
\hline
\end{tabular}
\end{center}
\vspace{-2em}
\end{table}

\section{Conclusion}   \label{conclusion}
In this letter,  we have proposed a novel class of sparse codebooks SCMA-aided NTNs with randomly distributed users characterized by Rician fading channels.  By looking into the PEP and the statistic order theory,  the BER performance with  randomly distributed users over Rician fading channels has been analyzed.   Considering the user fairness in the SCMA systems, we aim to minimize the  BER performance of the worst user.   Additionally,  the PAM constellation has been   employed to design the MC, where the codebooks are obtained by applying user-specific rotation and power factors. To minimize the number of parameters requiring optimization while preserving power diversity across users, an orthogonal resource allocation strategy has been further proposed. Simulation results demonstrated the superior performance of our proposed codebook over existing ones.

\ifCLASSOPTIONcaptionsoff
  \newpage
\fi

\bibliographystyle{IEEEtran} 
\bibliography{ref} 




\end{document}